\documentclass[a4paper,english,aps,manuscript,showpacs]{revtex4}
\usepackage[T1]{fontenc}
\usepackage[latin9]{inputenc}
\usepackage[active]{srcltx}
\usepackage{amstext}
\usepackage{amssymb}

\makeatletter

\@ifundefined{textcolor}{}
{%
 \definecolor{BLACK}{gray}{0}
 \definecolor{WHITE}{gray}{1}
 \definecolor{RED}{rgb}{1,0,0}
 \definecolor{GREEN}{rgb}{0,1,0}
 \definecolor{BLUE}{rgb}{0,0,1}
 \definecolor{CYAN}{cmyk}{1,0,0,0}
 \definecolor{MAGENTA}{cmyk}{0,1,0,0}
 \definecolor{YELLOW}{cmyk}{0,0,1,0}
}

\makeatother

\usepackage{babel}
\begin{document}

\title{Intermode-coupling modulation in the  fermion-boson model: heating
effects in the BCS regime}

\author{J. Plata}

\affiliation{Departamento de F\'{\i}sica, Universidad de La Laguna,\\
 La Laguna E38204, Tenerife, Spain.}
\begin{abstract}
Heating induced by an oscillating modulation of the interaction strength
in an atomic Fermion pair condensate is analyzed. The coupled fermion-boson
model, generalized by incorporating a time-dependent intermode coupling
through a magnetic Feshbach resonance, is applied. The dynamics is
analytically characterized in a perturbative scheme. The results account
for experimental findings which have uncovered a damped and delayed
response of the condensate to the modulation. The delay is due to
the variation of the quasiparticle energies and the subsequent relaxation
of the condensate. The detected damping results from the excitations
induced by a nonadiabatic modulation: for driving frequencies larger
than twice the pairing gap, quasiparticles are generated, and, consequently,
heating sets in. 
\end{abstract}


\maketitle

\section{Introduction }

The realization of the crossover from a molecular Bose-Einstein condensate
(BEC) to a Bardeen-Cooper-Schrieffer (BCS) superfluid of atom pairs
\cite{key-1,key-2,key-3,key-4,key-5,key-6,key-7,key-8,key-9,key-10,key-11,key-12}
has opened the way to a variety of experiments on fundamental effects
in quantum statistics and many-body physics. Essential to the versatility
of this scenario has been the possibility of varying the interaction
strength via a Feshbach resonance (FR), which has allowed the characterization
of the effects under controlled conditions. In parallel, the need
of explaining emergent phenomenology has brought about active theoretical
work in the field. Recent research has dealt with the role of thermal
fluctuations and the characterization of nonequilibrium situations
\cite{key-13}. In this line, here, we extend previous theoretical
work on the effects of an oscillating modulation of the interaction
strength in a two-component Fermi gas of atoms \cite{key-14}. Our
objective is to complete the analysis  of the experiments of Ref.
{[}15{]}. In them, a gas of ultracold $^{6}Li$ atoms was prepared
in the BCS regime through a magnetic FR, specifically, the (broad)
FR at 834 G between the two lowest hyperfine states. The application
of a sinusoidal modulation of the magnetic field was shown to lead
the condensate fraction to oscillate with the driving frequency. Moreover,
the oscillations were found to be damped and delayed with respect
to the modulation, the damping time being much longer than the driving
period. In previous work \cite{key-14}, the focus was put on the
mechanism responsible for the delay: the modulation was shown to drive
the system to an out-of-equilibrium situation, the deferred response
being rooted in the finite relaxation time of the condensate. Here,
we will concentrate on understanding the decay of the condensate fraction.
To this end, we set up a framework where, through a partial analytical
characterization of the dynamics, the origin of the damping processes
can be identified.  The decay of the oscillations will be linked to
heating induced by the generation  of quasiparticles. Nonadiabaticity,
(on the gap time scale), will appear as a crucial component of the
excitation process. Our scheme will allow us to trace the differential
aspects of the mechanisms responsible for heating and delay in the
system response. 

The outline of the paper is as follows. In Section II, we generalize
the coupled fermion-boson model \cite{key-16,key-17,key-18,key-19,key-20,key-21}
by incorporating a time-dependent intermode detuning, or, equivalently,
a time-dependent coupling. To deal with this variation of the basic
model, a perturbative scheme based on the Hartree-Fock-Bogoliubov
(HFB) description is developed. In Section III, the relaxation of
the condensate and the associated delayed response to the modulation
are tackled. The heating effects are evaluated in Section IV. In Section
V, our conclusions are summarized.

\section{The  fermion-boson model with a modulated coupling strength}

Our system consists of a gas of ultracold fermionic atoms with two
hyperfine states coupled to a molecular two-particle state through
a magnetic FR. To describe it, we apply the coupled fermion-boson
model \cite{key-16,key-18,key-19,key-20}. The grand-canonical Hamiltonian
reads 

\begin{eqnarray}
H-\mu N & = & \sum_{\mathbf{k},\sigma}\varepsilon_{\mathbf{k}}a_{\mathbf{k},\sigma}^{\dagger}a_{\mathbf{k},\sigma}+V_{int}\sum_{\mathbf{q,k,k^{\prime}}}a_{\frac{\mathbf{q}}{2}+\mathbf{k},\uparrow}^{\dagger}a_{\frac{\mathbf{q}}{2}-\mathbf{k},\downarrow}^{\dagger}a_{\frac{\mathbf{q}}{2}-\mathbf{k^{\prime}},\downarrow}a_{\frac{\mathbf{q}}{2}+\mathbf{k^{\prime}},\uparrow}+\nonumber \\
 &  & \sum_{\mathbf{q}}\left(\varepsilon_{\mathbf{q}}^{m}+\hbar\nu_{0}\right)b_{\mathbf{q}}^{\dagger}b_{\mathbf{q}}+g\sum_{\mathbf{q,k}}\left(b_{\mathbf{q}}a_{\frac{\mathbf{q}}{2}+\mathbf{k},\uparrow}^{\dagger}a_{\frac{\mathbf{q}}{2}-\mathbf{k},\downarrow}^{\dagger}+\textnormal{h.c.}\right)
\end{eqnarray}
where $\mu$ is the chemical potential, $N$ is the total number of
atoms, $a_{\mathbf{k},\sigma}^{\dagger}$ ($a_{\mathbf{k},\sigma}$)
denotes a fermionic creation (annihilation) operator of an atom with
momentum $\mathbf{k}$ and spin $\sigma$, ($\sigma\in\left\{ \uparrow,\downarrow\right\} $),
and $b_{\mathbf{q}}^{\dagger}$ ($b_{\mathbf{q}}$) is a bosonic operator
that creates (destroys) a molecule with momentum $\mathbf{q}$. It
is assumed that the two hyperfine states are equally populated. $\varepsilon_{\mathbf{k}}=\hbar^{2}\textrm{\ensuremath{\mathbf{\textrm{k}}^{2}}}/2m-\mu$
and $\varepsilon_{\mathbf{q}}^{m}=\hbar^{2}q^{2}/4m-2\mu$ are the
free dispersion relations for fermions and bosons, respectively. $V_{int}$(<$0$)
characterizes the binary attractive interaction potential between
fermions. Additionally, $g$ represents the FR coupling between the
closed and the open channel states, and $\nu_{0}$ is the detuning
of the boson resonance state from the collision continuum. (We stress
that the considered approach is actually a realization of a general
two-channel model to the particular context of Fermion-pair production.) 

Initially, the system is at equilibrium at a finite temperature $T$.
In that situation, a sinusoidal modulation of the detuning from the
FR is applied. Correspondingly, $\nu_{0}$ is replaced in Eq. (1)
by $\nu(t)=\nu_{0}+A\sin\omega_{p}t$. It is assumed that $V_{int}$,
which corresponds to the pairing interaction resulting from nonresonant
processes, is not affected by the applied driving field\emph{.} Through
the unitary transformation $U(t)=e^{i\frac{A}{\omega_{p}}\cos\omega_{p}t\left(\sum_{\mathbf{q}}b_{\mathbf{q}}^{\dagger}b_{\mathbf{q}}+\frac{1}{2}\sum_{\mathbf{k},\sigma}a_{\mathbf{k},\sigma}^{\dagger}a_{\mathbf{k},\sigma}\right)}$,
the Hamiltonian, transformed as $H^{\prime}=U^{\dagger}HU-i\hbar U^{\dagger}\dot{U}$,
is converted into 

\begin{eqnarray}
H^{\prime}-\mu N & = & \sum_{\mathbf{k},\sigma}\left(\varepsilon_{\mathbf{k}}-\hbar\frac{A}{2}\sin(\omega_{p}t)\right)a_{\mathbf{k},\sigma}^{\dagger}a_{\mathbf{k},\sigma}+V_{int}\sum_{\mathbf{q,k,k^{\prime}}}a_{\frac{\mathbf{q}}{2}+\mathbf{k},\uparrow}^{\dagger}a_{\frac{\mathbf{q}}{2}-\mathbf{k},\downarrow}^{\dagger}a_{\frac{\mathbf{q}}{2}-\mathbf{k^{\prime}},\downarrow}a_{\frac{\mathbf{q}}{2}+\mathbf{k^{\prime}},\uparrow}+\nonumber \\
 &  & \sum_{\mathbf{q}}(\varepsilon_{\mathbf{q}}^{m}+\hbar\nu_{0})b_{\mathbf{q}}^{\dagger}b_{\mathbf{q}}+\left(g\sum_{\mathbf{q,k}}b_{\mathbf{q}}a_{\frac{\mathbf{q}}{2}+\mathbf{k},\uparrow}^{\dagger}a_{\frac{\mathbf{q}}{2}-\mathbf{k},\downarrow}^{\dagger}+\textnormal{h.c.}\right).
\end{eqnarray}
(In obtaining the above expression we have made use of the partial
result $U^{\dagger}HU=H$, which derives from the commutation relation
of $H$ with the total number of fermions and the cancellation of
the introduced time dependence in the interaction term.) (In Ref.
{[}14{]}, an alternative approach was implemented by transferring
the time variation in $\nu$ to the intermode coupling via a different
unitary transformation. Note, that, as shown in Ref. {[}14{]}, the
time dependence of the coupling term prevents the one-mode reduction,
applicable to the undriven dynamics for broad resonances.) An approximate
description of the dynamics resulting from the modulation can be obtained
through the following perturbative scheme. The complete Hamiltonian
is split as $H^{\prime}-\mu N\simeq H_{0}+H_{per}$, where the unperturbed
Hamiltonian has the form given by Eq. (1), i.e., $H_{0}=H-\mu N$;
and, the perturbation reads 

\[
H_{per}=-\frac{\hbar}{2}A\sin(\omega_{p}t)\sum_{\mathbf{k,\sigma}}a_{\mathbf{k},\sigma}^{\dagger}a_{\mathbf{k},\sigma}.
\]
(We consider that the modulation amplitude is sufficiently small for
the perturbative scheme to be valid. It is assumed that the system,
initially in the BCS side, stays in that regime during the whole process.
Hence, the BEC side is not reached and neither is attained the unitary
limit. Later on, we will precisely define the range of applicability
of our approach.)

\subsection{The zero-order Hamiltonian}

To describe the unperturbed system, we follow the standard HFB approach
\cite{key-20,key-18}. Accordingly, we introduce first three mean
fields: $n_{0}\equiv\sum_{\mathbf{k}}\left\langle a_{\mathbf{k},\sigma}^{\dagger}a_{\mathbf{k},\sigma}\right\rangle $
for the spin density, $\Delta_{0}\equiv\left|V_{int}\right|\sum_{\mathbf{k}}\left\langle a_{-\mathbf{k},\downarrow}a_{\mathbf{k},\uparrow}\right\rangle $
for the pairing field, and $\phi_{m,0}\equiv\left\langle b_{\mathbf{q=0}}\right\rangle $
for the boson field. (We take $\mathbf{q}=\mathbf{0}$ as we focus
on the condensed molecular field.) Through the incorporation of those
mean fields, the zero-order Hamiltonian, which describes the unmodulated
system, is rewritten in the form 
\[
H_{0}=\sum_{\mathbf{k},\sigma}V_{k}a_{\mathbf{k},\sigma}^{\dagger}a_{\mathbf{k},\sigma}-\sum_{\mathbf{k}}(\tilde{\Delta}_{0}a_{\mathbf{k},\uparrow}^{\dagger}a_{-\mathbf{k},\downarrow}^{\dagger}+\textnormal{h.c.}),
\]
which corresponds to an effective BCS model with mode energy $V_{k}\equiv\varepsilon_{\mathbf{k}}+V_{int}n_{0}$
and gap $\tilde{\Delta}_{0}\equiv\Delta_{0}-g\phi_{m,0}$. The mean-field
description includes also the equation for the evolution of the boson
mode, namely,

\textbf{
\begin{equation}
i\hbar\frac{d\phi_{m,0}}{dt}=(\nu_{0}-2\mu)\phi_{m,0}+\frac{g}{\left|V_{int}\right|}\Delta_{0}.
\end{equation}
}

$H_{0}$ is straightforwardly diagonalized. By applying the Bogoliubov
transformation (BT) defined by the fermionic operators $c_{\mathbf{k},\uparrow}=\cos\theta_{k}a_{\mathbf{k},\uparrow}-\sin\theta_{k}a_{-\mathbf{k},\downarrow}^{\dagger}$
and $c_{-\mathbf{\mathbf{k},\downarrow}}^{\dagger}=\sin\theta_{k}a_{\mathbf{k},\uparrow}+\cos\theta_{k}a_{-\mathbf{k},\downarrow}^{\dagger}$,
where $\theta_{k}$ is given by $\tan(2\theta_{k})=\left|\tilde{\Delta}_{0}\right|/V_{k}$
\cite{key-18,key-20}, we find 
\[
H_{0}=\sum_{\mathbf{k}}E_{k,0}(c_{\mathbf{k},\uparrow}^{\dagger}c_{\mathbf{k},\uparrow}+c_{\mathbf{\mathbf{k},\downarrow}}^{\dagger}c_{\mathbf{\mathbf{k},\downarrow}})+\textrm{constant}.
\]
The operator $c_{\mathbf{k},\uparrow}^{\dagger}$ ($c_{\mathbf{k},\uparrow}$)
creates (annihilates) a quasi-particle excitation with momentum $\mathbf{k}$
and spin $\uparrow$. The associated excitation energies are

\begin{equation}
E_{k,0}=\sqrt{V_{k}^{2}+\tilde{\Delta}_{0}^{2}}=\sqrt{(\hbar^{2}k^{2}/2m-\mu+V_{int}n_{0})^{2}+\tilde{\Delta}_{0}^{2}}
\end{equation}
The BCS state $\left|\Psi_{BCS}\right\rangle $ is the effective vacuum
state of this Hamiltonian, i.e., $c_{\mathbf{k},\uparrow}\left|\Psi_{BCS}\right\rangle =c_{\mathbf{\mathbf{k},\downarrow}}\left|\Psi_{BCS}\right\rangle =0$;
in the previous representation, it is given by $\left|\Psi_{BCS}\right\rangle =\prod_{\mathbf{k}}\left(\cos\theta_{k}+\sin\theta_{k}a_{\mathbf{k},\uparrow}^{\dagger}a_{-\mathbf{k},\downarrow}^{\dagger}\right)\left|0\right\rangle $
\cite{key-12}. Note that the excitation gap $\tilde{\Delta}_{0}$
combines the mean pairing field $\Delta_{0}$ and the equilibrium
molecular field $\phi_{m,0}$. $\Delta_{0}$ is obtained from the
BCS equation 
\begin{equation}
\Delta_{0}=\frac{\left|V_{int}\right|}{2}\sum_{\mathbf{k}}[2f_{k}^{eq}(0)-1]\sin(2\theta_{k}),
\end{equation}
where $\left\{ f_{k}^{eq}(0)\right\} $ are the initial populations
of the quasiparticle states, which are given by the Fermi distribution
function, i.e., $f_{k}^{eq}(0)=1/(1+e^{E_{k,0}/k_{B}T})$, since thermal
equilibrium is assumed for the system before the application of the
magnetic modulation. Moreover, $\phi_{m,0}$ is obtained as the stationary
solution to Eq. (3), namely, 
\begin{equation}
\phi_{m,0}=\frac{g\Delta_{0}}{\left|V_{int}\right|(2\mu-\nu_{0})}.
\end{equation}
(Note that a selfconsistent procedure is required to obtain $\Delta_{0}$
and $\phi_{m,0}$.)

\subsection{The perturbation}

We turn now to analyze the effect of the perturbation. The introduction
of $H_{per}$ in the HFB approach implies dealing with changes in
the mean fields, which now become $n(t)=n_{0}+\delta n(t)$, $\Delta(t)=\Delta_{0}+\delta\Delta(t)$,
and $\phi_{m}(t)=\phi_{m,0}+\delta\phi_{m}(t)$. Our objective is
solving for the perturbation-induced increments of those fields. In
particular, we will focus on explaining the damping and delay of $\delta\Delta(t)$
observed in the experiments. Let us see that a first-order approximation
to that behavior can be obtained simply by incorporating $H_{per}$
into the HFB approach defined by the unperturbed mean fields. Specifically,
we apply the previously defined BT to the perturbation Hamiltonian,
which, as a result, is written as $H_{per}=H_{evar}+H_{coup}$, where 

\begin{eqnarray}
H_{evar} & =-\frac{\hbar A}{2}\sin(\omega_{p}t) & \sum_{\mathbf{k}}\cos(2\theta_{k})\left(c_{\mathbf{k},\uparrow}^{\dagger}c_{\mathbf{k},\uparrow}+c_{\mathbf{\mathbf{k},\downarrow}}^{\dagger}c_{\mathbf{\mathbf{k},\downarrow}}\right),
\end{eqnarray}

\begin{eqnarray}
H_{coup} & =-\frac{\hbar A}{2}\sin(\omega_{p}t) & \sum_{\mathbf{k}}\sin(2\theta_{k})\left(c_{\mathbf{k},\uparrow}^{\dagger}c_{-\mathbf{k},\downarrow}^{\dagger}+\textnormal{h.c.}\right).
\end{eqnarray}
 From the forms of $H_{evar}$ and $H_{coup}$, two preliminary general
conclusions on the effect of the magnetic modulation can be drawn.
First, $H_{evar}$ leads to a time variation of the quasiparticle
energies, which become $E_{k}(t)=E_{k,0}+\delta E_{k}(t)$, where

\begin{equation}
\delta E_{k}(t)=-\frac{\hbar A}{2}\sin(\omega_{p}t)\cos(2\theta_{k})
\end{equation}
In the next section, we will see that the disequilibrium induced by
this term and the subsequent relaxation of the condensate are at the
origin the detected delayed response of the system to the driving.
Second, $H_{coup}$ represents modulation-induced interactions between
the vacuum state and a doubly-excited state. Importantly, these coupling
terms, which oscillate with the external frequency $\omega_{p}$,
are relevant only when they can induce an effective resonance between
the BCS state and the two-excitation configuration, i.e., only when
$\hbar\omega_{p}\geq2\tilde{\Delta}_{0}$. The resulting heating effects
will be analyzed in Section IV.

\section{The effect of the quasiparticle-energy variation}

In the regime defined by $\hbar\omega_{p}<2\tilde{\Delta}_{0}$, the
interaction terms given by $H_{coup}$ can be discarded, and, consequently,
the perturbation Hamiltonian $H_{per}$ can be approximated as $H_{evar}=\sum_{\mathbf{k}}\delta E_{k}(c_{\mathbf{k},\uparrow}^{\dagger}c_{\mathbf{k},\uparrow}+c_{\mathbf{\mathbf{k},\downarrow}}^{\dagger}c_{\mathbf{\mathbf{k},\downarrow}})$.
Hence, the driven Hamiltonian is still diagonal in the representation
of the quasiparticle states of the unmodulated system. We will see
that, although $H_{evar}$ simply leads to the variation of the quasi-particle
energies, the consequent effect on the gap dynamics can be quite complex;
in fact, its analysis will require the generalization of our model.
(Here, we will follow a treatment alternative to that presented in
Ref. {[}14{]}, which will allow us to simplify the characterization
of the basic physics of the delay.)  

The perturbation forces the system out of equilibrium: the initial
thermal populations associated with the unmodulated energies do not
fit the Fermi distribution $f_{k}^{eq}(t)=1/(1+e^{E_{k}(t)/k_{B}T})$
for the actual (time-varying) energies. Indeed, the gap equation now
reads 

\begin{eqnarray}
\Delta(t) & =\frac{\left|V_{int}\right|}{2} & \sum_{\mathbf{k}}\left[2f_{k}(t)-1\right]\sin(2\theta_{k}),
\end{eqnarray}
where $\left\{ f_{k}(t)\right\} $ are the (changing) populations.
To describe the dynamics, we must deal with the relaxation of the
populations towards equilibrium, which implies extending the current
Hamiltonian description. Note that a selfconsistent approach is needed.
The evolving $\left\{ E_{k}(t)\right\} $ affect the relaxation of
the populations $\left\{ f_{k}(t)\right\} $ by modifying the equilibrium
distribution $\left\{ f_{k}^{eq}(t)\right\} $. In turn, the variation
of $\Delta(t)$ changes the global Hamiltonian, and, in particular,
can alter the quasiparticle energies. As previously stated, in the
simplified description considered here, we neglect corrections to
the quasi-particle energies due to changes in the mean fields: the
form given by Eq. (9) is assumed to permanently apply. (See Ref. {[}14{]}
for an analysis of higher-order effects.) We will see that this simplification
retains the system components responsible for the emergence of the
features observed in the experiments.

\subsection{The relaxation mechanism}

The mechanism for thermalization can be assumed to be based on collisions
between excited particles. Here, instead of tackling a detailed analysis
of the dependence of the relaxation on the system characteristics,
we will focus on general aspects of its role in the condensate dynamics.
Accordingly, we consider that the evolution of the populations is
governed by the generic equation \cite{key-22}

\begin{eqnarray}
\frac{df_{k}}{dt} & = & -\frac{1}{\tau_{f}}\left[f_{k}(t)-f_{k}^{eq}(t)\right],
\end{eqnarray}
where $1/\tau_{f}$ represents the effective thermalization rate.
No restrictions on the magnitude of $\tau_{f}$ are assumed. A similar
relaxation mechanism was considered in Ref. {[}22{]} in the context
of nonequilibrium superconductivity. Central to this mechanism is
the idea that the relaxation is activated by the distance from the
actual populations to those corresponding to the equilibrium, which
are, in turn, changing as the quasiparticles energies are being modified
by the driving. We have assumed a first (compact) form of characterizing
that process with the introduction of the effective thermalization
rate. Note that, since the system is continuously forced out of equilibrium
by the driving field, i.e., the $f_{k}^{eq}(t)$ are permanently changing,
the relaxation mechanism is always activated. The nondirect following
to the modulation observed in the experiments can be anticipated to
be rooted in finite values of $\tau_{f}$.

Eq. (11) is an inhomogeneous linear differential equation, which is
exactly solved to give 

\[
f_{k}(t)=e^{-t/\tau_{f}}\left(f_{k}(0)-\frac{1}{\tau_{f}}\int_{0}^{t}e^{t^{\prime}/\tau_{f}}f_{k}^{eq}(t^{\prime})dt^{\prime}\right).
\]
Furthermore, through integration by parts, we find 

\begin{equation}
f_{k}(t)=e^{-t/\tau_{f}}\left(f_{k}(0)-f_{k}^{eq}(0)\right)+f_{k}^{eq}(t)-e^{-t/\tau_{f}}\int_{0}^{t}e^{t^{\prime}/\tau_{f}}\frac{df_{k}^{eq}}{dt}(t^{\prime})dt^{\prime}.
\end{equation}
This expression is simplified by taking $f_{k}(0)=f_{k}^{eq}(0)$,
since the system is at equilibrium at $t=0$. By combining Eqs. (10)
and (12), we obtain the following integral-differential equation for
the order parameter

\emph{
\begin{eqnarray}
\Delta(t) & =\frac{\left|V_{int}\right|}{2} & \sum_{\mathbf{k}}\left[2\left(f_{k}^{eq}(t)-\int_{-\infty}^{t}e^{-(t-t^{\prime})/\tau_{f}}\frac{df_{k}^{eq}}{dt}(t^{\prime})dt^{\prime}\right)-1\right]\sin(2\theta_{k}).
\end{eqnarray}
}We deal now with particular regimes where we can go further in the
analytical characterization of the gap evolution, and, consequently,
in the identification of the delay time.

\subsection{The response to the modulation at small departure from equilibrium}

Eq. (13) simplifies considerably in the regime defined by $E_{k}\sim\Delta\ll T\approx T_{c}$,
($k_{B}=1$). ($T_{c}$ is the temperature for the BCS transition.)
In this range, the approximations $f_{k}^{eq}(t)\simeq f_{k}^{eq}(0)+\frac{df_{k}^{eq}}{dE_{k}}\delta E_{k}(t)$
and $\frac{df_{k}^{eq}}{dE_{k}}\simeq-\frac{1}{4T_{c}}$ can be made
\cite{key-22}. In turn, we can write $\frac{df_{k}^{eq}}{dt}\simeq-\frac{1}{4T_{c}}\frac{dE_{k}}{dt}=\frac{\hbar A\omega_{p}}{8T_{c}}\cos(\omega_{p}t)\cos(2\theta_{k})$.
Through the incorporation of these approximations into Eq. (12), and,
subsequent integration, we obtain for the populations 

\begin{eqnarray}
f_{k}(t) & =f_{k}^{eq}(0) & +\frac{\hbar A}{8T_{c}}\frac{\cos(2\theta_{k})}{1+(\omega_{p}\tau_{f})^{2}}\left[\sin(\omega_{p}t)-\omega_{p}\tau_{f}\left(\cos(\omega_{p}t)-e^{-t/\tau_{f}}\right)\right].
\end{eqnarray}
Then, combining this equation with Eq. (13), we find 

\begin{eqnarray}
\frac{\Delta(t)}{\Delta_{0}} & =1 & +C\left[e^{-t/\tau_{f}}\sin\varphi+\sin(\omega_{p}t-\varphi)\right],
\end{eqnarray}
where

\begin{equation}
C=\frac{\left|V_{int}\right|}{\Delta_{0}}\frac{\hbar A}{8T_{c}\sqrt{1+(\omega_{p}\tau_{f})^{2}}}\sum_{\mathbf{k}}\sin(2\theta_{k})\cos(2\theta_{k}),
\end{equation}
and

\[
\varphi=\arctan(\omega_{p}\tau_{f}).
\]
Some implications of these results must be stressed:

(i) The gap evolution incorporates a transitory decay with characteristic
time $\tau_{f}$ and a secular oscillatory behavior with frequency
$\omega_{p}$. The external field is not instantaneously followed:
associated with the phase shift $\varphi$, there is a delay time
given by $\tau_{D}=\frac{\varphi}{\omega_{p}}=\tau_{f}\left[1+\mathcal{O}\left((\omega_{p}\tau_{f})^{2}\right)\right]$,
which can be interpreted as the condensate relaxation time. No changes
in the delay are observed at different cycles of the field modulation
in agreement with the experimental results. Furthermore, the detected
invariance of the delay with the external frequency can be understood
as associated with the small magnitude of the correction $\mathcal{O}\left((\omega_{p}\tau_{f})^{2}\right)$
for the experimental conditions. The complex character of the driving
mechanism is apparent in the obtained expression for the amplitude
of the oscillatory term, which combines external-field parameters
and characteristics of the unperturbed system. (The factor $\sum_{\mathbf{k}}\sin(2\theta_{k})\cos(2\theta_{k})$
present in Eq. (16) can be standardly evaluated \cite{key-22,key-23}.)
It is worth emphasizing that the effect of the perturbation scales
with the factor $A/\sqrt{1+(\omega_{p}\tau_{f})^{2}}$. 

(ii) Outside the considered regime, the intricate interdependence
of the gap and the populations can imply a complex nonlinear contribution
of the populations to the gap relaxation \cite{key-22}. Hence, one
can expect that, in a general  regime, the delay time can significantly
differ from $\tau_{f}$.

(iii) The consistency of our approach can be tested by analyzing the
limits of small and large relaxation time $\tau_{f}$. When $\tau_{f}$
is much smaller than any other characteristic time in the process,
in particular, than the driving period, we find that $\varphi\rightarrow0$
\cite{key-24}. Then, there is no delay between the gap evolution
and the external field. The predictions of the adiabatic approximation
are consistently reproduced: for a sudden relaxation, the populations
follow adiabatically, (on the relaxation time scale), the equilibrium
values $\left\{ f_{k}^{eq}(t)\right\} $ associated with the time-dependent
energies. The associated gap dynamics becomes ``trivial'': the evolution
corresponds to a sequence of equilibrium states where time enters
as a parameter.  On the other hand, for a very large $\tau_{f}$,
we obtain $C\rightarrow0$, and, therefore, $\Delta(t)=\Delta_{0}$:
we trivially recover that there is no change in the gap for times
much smaller than the characteristic time for the evolution of the
populations $\tau_{f}$.

\section{Heating effects}

For $\hbar\omega_{p}\geq2\tilde{\Delta}_{0}$, the term $H_{coup}$
in the perturbation becomes relevant: the magnetic field can then
induce an effective resonance between the fundamental state and a
doubly-excited state. (See Refs. {[}25{]} and {[}26{]} for related
work.) To analyze the resulting transition, we rewrite the coupling
Hamiltonian as 

\begin{eqnarray*}
H_{coup} & =\hat{W}\sin(\omega_{p}t) & ,
\end{eqnarray*}
where

\[
\hat{W}=-\frac{\hbar A}{2}\sum_{\mathbf{k}}\sin(2\theta_{k})\left(c_{\mathbf{k},\uparrow}^{\dagger}c_{-\mathbf{k},\downarrow}^{\dagger}+\textnormal{h.c.}\right).
\]
Note that, because of the dependence of the factor $\sin(2\theta_{k})=\left|\tilde{\Delta}_{0}\right|/E_{k,0}$
on the quasiparticle energy, the coupling is less effective as the
energy grows. (The opposite occurs in the term $H_{evar}$, {[}see
Eq. (7){]}.) The combination of this characteristic with the form
of the density of states will be shown to determine prominent features
of the system response. The transfer from the BCS state $\left|\Psi_{BCS}\right\rangle $
to the doubly-excited state $\left|f\right\rangle =c_{\mathbf{k},\uparrow}^{\dagger}c_{-\mathbf{k},\downarrow}^{\dagger}\left|\Psi_{BCS}\right\rangle $
of the continuum of quasi-particle states can be evaluated applying
Fermi's Golden Rule. (The reverse process, i.e., the decay of doubly-excited
states induced by $H_{coup}$, can be neglected: given the range of
temperatures considered, the population of excited states is always
much smaller than that of the fundamental state. Also relevant to
the lack of symmetry in the reverse transition is the continuum structure
of the excited states.) As corresponds to a sinusoidal perturbation,
we have for the transition rate: 

\begin{eqnarray}
\gamma(\omega_{p}) & = & \frac{\pi}{2\hbar}\sum_{f}\left|\left\langle f\left|\hat{W}\right|\Psi_{BCS}\right\rangle \right|^{2}\delta(E_{f}-E_{BCS}-\hbar\omega_{p})\nonumber \\
 & = & \frac{\pi\hbar}{8}A^{2}\sum_{\mathbf{k}}\sin^{2}(2\theta_{k})\delta(2E_{k,0}-\hbar\omega_{p}).
\end{eqnarray}
The sum in $\mathbf{k}$ is standardly converted into an integral:
$\sum_{\mathbf{k}}\rightarrow\frac{\mathcal{V}}{(2\pi)^{3}}\int d\mathbf{k}$
\cite{key-27}. {[}Here, $\mathcal{V}$ denotes the quantization volume,
which disappears in the final expression as a scaling with $\mathcal{V}^{-1/2}$
is incorporated into the definition of the operators introduced in
Eq. (1){]}. The integral is evaluated by changing to the variable
$E_{k,0}$. Indeed, incorporating the density of states obtained from
the dispersion relation given by Eq. (4), the transition rate is found
to be given by 

\begin{eqnarray}
\gamma(\omega_{p}) & = & \frac{1}{2\pi}\left(\frac{m}{2\hbar^{2}}\right)^{3/2}\tilde{\Delta}_{0}^{2}\tilde{A}^{2}\frac{1}{\omega_{p}}\left[\sqrt{(\hbar\omega_{p}/2)^{2}-\tilde{\Delta}_{0}^{2}}+\mu-V_{int}n_{0}\right]^{1/2}\times\nonumber \\
 &  & \left[(\hbar\omega_{p}/2)^{2}-\tilde{\Delta}_{0}^{2}\right]^{-1/2}\Theta(\hbar\omega_{p}-2\tilde{\Delta}_{0}),
\end{eqnarray}
where $\tilde{A}=A\mathcal{V}^{1/2}$, and $\Theta(x)$ is the Heaviside
step function: the excitation process is activated only for frequencies
equal or larger than the threshold value $\hbar\omega_{p}^{th}=2\tilde{\Delta}_{0}$.
Note that the divergence at the threshold is a consequence of the
singularity of the density of states in the BCS model at $E_{k,0}=\tilde{\Delta}_{0}$.
It is important to take into account that this failure of the perturbative
scheme does not invalidate the identification of the physical mechanism
responsible for heating: the fast decrease of the transition rate
with $\omega_{p}$ allows the applicability of the used approach sufficiently
far from the threshold. 

The transfer of population from the fundamental state to the excited
states implies the decrease of $\Delta(t)$. This is apparent from
Eq. (10): although the total population is conserved in the transition,
$\Delta(t)$ diminishes since the factor $\sin(2\theta_{k})$ decreases
as $E_{k,0}$ grows. Since the excitation is permanently activated
by the driving, the resulting damping process is continuous, in agreement
with the character of the detected decay of the condensate fraction.
The global picture of the gap dynamics that emerges from combining
this effect with the delayed oscillation analyzed in Sec. III corresponds
to the behavior detected in the experiments. Note that, in the applied
approach, the mechanisms for delay and damping can be considered to
work in parallel. In this sense, it is worth pointing out that the
term of population gain for the excited states that, because of heating,
should be added to Eq. (11) is irrelevant given the small contribution
of those high-level populations to the gap dynamics.

The applicability of the perturbative approach can be assessed from
the analysis of the dependence of the system output on the modulation
parameters. Both, the damping coefficient $\gamma$ and the amplitude
of the oscillatory component $C$, given by Eq. (16), diminish for
decreasing modulation amplitudes and growing frequencies.  The observation
of oscillatory behavior along with damping requires working with decay
times larger than the driving period. Given the dependence of $\gamma$
on $A^{2}$, and the requirement $\hbar\omega_{p}^{th}\geq2\tilde{\Delta}_{0}$
for the emergence of damping, that situation can occur for sufficiently
small modulation amplitudes and large frequencies. Then, it seems
possible to reproduce that situation in a range of parameters where
the applicability of the perturbative scheme can be guaranteed. Technical
details of the measurement of the condensate fraction, which is the
magnitude reproduced by our model, can be found in Ref. {[}15{]}. 

We have assumed that the trapping conditions implemented in the referred
experiments do not crucially affect the main characteristics of the
observed features. Indeed, as our uniform description qualitatively
reproduces the experimental results, the robustness of the identified
physical mechanisms against spatial non-uniformities can be conjectured.
For a smooth external potential $U(\vec{r})$, which corresponds to
the practical conditions, a local-density approximation can be applied
to generalize the uniform picture. Accordingly, trapping can be incorporated
in the previous framework by replacing the chemical potential as $\mu\rightarrow\mu(\vec{r})=\mu-U(\vec{r})$
\cite{key-27}. The consequent use of local fields implies less compact
results  but does not affect the basic physics underlying the observed
features. One of the effects of nonuniformity on heating seems evident:
the local fields $\Delta_{0}(\vec{r})$, $\phi_{m,0}(\vec{r})$, $n_{0}(\vec{r})$,
and, in turn, $\tilde{\Delta}_{0}(\vec{r})$, can be expected to smoothen
the sharp threshold  for the onset of excitations. Although the previously
derived expressions can be modified by the averaging over the distribution
along the trap, the former basic picture still applies.

\section{Concluding remarks}

The considered variation of the coupled fermion-boson model has been
shown to give  useful clues to understanding the dynamics of atomic
fermion pairs with modulated interaction strength. The perturbative
scheme set up from the basic BCS approach has allowed isolating the
role of the different elements of the system. The emergence of specific
dynamical features has been found to depend on the time scales of
the system components: the relative magnitudes of the driving period,
the inverse gap frequency, and the relaxation time determine the characteristics
of the system response. The interest of further experimental work
on specific aspects of the modulation scheme is evident. Particularly
valuable can be the experimental characterization of the system response
for driving frequencies slightly larger than the threshold, where
the perturbative approach fails. Also interesting can be checking
the inhibition of damping for frequencies smaller than $2\tilde{\Delta}_{0}$. 

It is of interest to mention recent related works on alternative approaches
to the study of similar systems. In this sense, it is pertinent to
point out the advances in the characterization of the system parameters
which are reported in Ref. {[}31{]} and the applications of analogue
methodology to p-wave interacting Fermi gases \cite{key-32}. Also
valuable is to establish a parallelism between the considered scenario
and similar modulation techniques applied in bosonic systems. (Ref.
{[}33{]} presents interesting experimental findings on the production
of ultracold molecules via a sinusoidal modulation of the magnetic
field. Subsequent theoretical analysis was presented in Ref. {[}34{]}.) 

Finally, it is worth pointing out that the applicability of the study
is not restricted to the field of ultracold atomic gases. The central
issue in the analysis, namely, the gap dynamics \cite{key-28}, in
particular, the effect of changes in the populations on the evolution
of the condensate fraction, is relevant to topics ranging from nonequilibrium
superconductivity \cite{key-22,key-29} to quenched dynamics in superfluid
$^{3}He$ \cite{key-30}. In this sense, the analytical character
of the study can be particularly useful given the difficulty of dealing
with out-of-equilibrium situations in those contexts.

\end{document}